\documentclass[a4paper,twoside]{article}
\usepackage[table]{xcolor}
\usepackage{booktabs}
\usepackage{graphicx}
\usepackage{epsfig}
\usepackage{subcaption}
\usepackage{calc}
\usepackage{amssymb}
\usepackage{amstext}
\usepackage{amsmath}
\usepackage{amsthm}
\usepackage{multicol}
\usepackage{pslatex}
\usepackage{apalike}
\usepackage{tcolorbox}
\usepackage{SCITEPRESS}     

\begin{document}

\title{Challenges Women in Software Engineering Leadership Roles Face: A Qualitative Study}

\author{\authorname{Karina Kohl\sup{1}\orcidAuthor{0000-0000-0000-0000} and Rafael Prikladnicki\sup{1}\orcidAuthor{0000-0000-0000-0000}}
\affiliation{\sup{1}School of Technology, PUCRS, Porto Alegre, Brazil}
\email{karina.kohl@edu.pucrs.br, rafael.prikladnicki@pucrs.br}
}


\abstract{Software engineering is not only about technical solutions. To a large extent, it is also concerned with organizational issues, project management, and human behavior. There are serious gender issues that can severely limit the participation of women in science and engineering careers. It is claimed that women lead differently than men and are more collaboration-oriented, communicative, and less aggressive than their male counterparts. However, when talking with women in technology companies' leadership roles, a list of problems women face grows fast. We invite women in software engineering management roles to answer the questions from an empathy map canvas. We used thematic analysis for coding the answers and group the codes into themes. From the analysis, we identified seven themes that support us to list the main challenges they face in their careers.}

\onecolumn \maketitle \normalsize \setcounter{footnote}{0} \vfill

\section{\uppercase{Introduction}}
\label{sec:introduction}

Software engineering is not only about technical solutions. It is, to a large extent, also concerned with organizational issues, project management, and human behavior \cite{Wohlin2003}. 

Diversity is being discussed intensively by different knowledge areas of society, and these discussions in Software Engineering are increasing as well. Different people form software development teams, and lately, it is being discussed that we have underrepresented groups as, for instance, gender, ethnic, cultural, and others. Page \cite{Page2007} defines cognitive diversity as the differences in how we interpret, reason, and solve, how we think and, identity diversity is determined by affiliation with a social group as gender, culture, ethnicity, religion, sexual orientation, etc. Cognitive diversity is linked to better outcomes in two main types of tasks: problem-solving and prediction. Identity diversity creates collective benefits when connects to cognitive diversity and connects these diverse talents to relevant problems.

Gender diversity often refers to an equitable or fair representation of people of different genders. It commonly refers to an equal ratio of men and women but may also include people of non-binary genders \cite{Sytsma2006}. Non-binary is a spectrum of gender identities that are not exclusively masculine or feminine, outside the gender binary \cite{usher2006}.

Page \cite{Page2007} says identity attributes cause us to construct different sets of cognitive tools. Identity differences lead to experiential differences that create tool differences. We can see this in the context of gender differences. Because we treat men and women differently, we provide them different experiences. As a result, they learn to think about situations differently. The effects of identity on experience and opportunities are hard to measure. Ideally, society would not discriminate based on identity characteristics. But even if society did not, policies that encourage or mandate identity blindness could not immediately overcome the residue of past biases.

Frize \cite{Frize2005} says there are serious gender issues that can severely limit women's participation in science and engineering careers, which are similar in many parts of the world. One main obstacle to women's retention or participation is that women's contributions and abilities are less valued than men's, and women are generally ignored in mainstream history. A systemic bias against women arose and was perpetuated for thousands of years by philosophers and thinkers. Aristotle's (384-322 BC) writings make his position clear: "\textit{The female is as it were a deformed male. The male is by nature fitter for the command than the female... The justice of a man and a woman is not the same; the courage of a man is shown in commanding a woman obeying. In men, qualities or capacities are found in their perfection, whereas women are less balanced, more easily moved to tears, more jealous; she is also more false of speech [and] more deceptive.}" Plato (429-347 BC), however, argued that women, like men, could rule and that those who showed such talents ought to be given access to education. If they are to become rulers, women must have equal access to education and training as men do. Plato advocated education for women so that they could participate equally in society \cite{Frize2005}. 

Jetter et al. \cite{Jetter2013} mention that leadership style, the "\textit{manner and approach of providing direction, implementing plans, and motivating people }," has a significant impact on team performance and the achievement of organizational goals. A participative leadership style was positively related to highly functional teams and fostered team innovation. Gender is a factor that has been investigated in leadership style studies. It is sometimes claimed that women lead differently than men and are more collaboration-oriented, communicative, and less aggressive than their male counterparts. 

Jetter and Walker \cite{Jetter2017} say it has almost become a stylized fact that, on average, women are more likely to avoid competition, under-perform in competitive environments, and exhibit higher risk aversion than men. Persistent social phenomena, such as the gender wage gap or the under-representation of women in highly competitive occupations and job positions, have been linked to such observations. One prominent hypothesis to explain this phenomenon relates to the idea that the gender of one's opposition could influence competitive behavior. More generally, people may behave differently when competing against adversaries from the opposite sex. If true, this would imply wide-ranging consequences in a number of settings. For instance, numerous work environments are characterized by persistent under-representation of one gender. Women are especially under-represented in jobs that are generally associated with high-pressure environments and large stakes, such as financial management (the share of females at Wall Street remains at approximately 10 percent) or CEO positions in the US (2.5 percent). Other areas with low female employee shares include IT- and math-related occupations, where women usually occupy less than 20 percent of positions.

Eagly and Carli \cite{EAGLY2003} say any female advantage in leadership style might be offset by disadvantage that flows from prejudice and discrimination directed against women as leaders. Prejudice consists of an unfair evaluation of a group of people based on stereotypical judgments of the group rather than its members' behavior or qualifications. When people hold stereotypes about a group, they expect that group to possess characteristics and exhibit behavior consistent with those stereotypes. They also say, consistent with role incongruity theory, stereotype research reveals that people do consider men to be more agentic than women and women to be more communal than men. Also, the communal qualities that people associate with women, such as warmth and selflessness, diverge from the agentic qualities, such as assertiveness and instrumentality, that people perceive as characteristics of successful leaders. In contrast, the predominantly agentic qualities that people associate with men are similar to the qualities perceived to be needed for success in high-status occupations, including most managerial occupations.

Novielli and Serebrenik \cite{Novielli2019} said that the interest in the power of emotions stimulated efforts to study the link between emotions and developers' productivity and understand the triggers for emotions during software development and related activities, and assess the impact of emotions on the developers' well-being. They say that a team manager or the scrum master can benefit from understanding the developers' emotions. Such an understanding can, for example, inform retrospective analysis that considers emotions for identifying and correcting uneven task distribution or for providing just-in-time support to a team member stuck in dealing with a programming task.

Considering the context of women in software engineering leadership roles, we used the empathy map canvas technique applied in the format of a survey to understand three women's perceptions and emotions in different management roles in a technology company (project/product managers, people managers, scrum masters, etc.). Our main goal is to answer the following research question:

\begin{tcolorbox}
\textit{RQ. What are the challenges women in Software Engineering leadership/management roles face?}   
\end{tcolorbox}

The rest of the paper proceeds as follows. Section \ref{sec:empathy} presents background about Empathy.  Section \ref{sec:design}  summarizes  the methodology we use in this study.  Section  \ref{sec:discussion}  presents the results we found and discussions.  Section  \ref{sec:conclusion} concludes the paper.

\section{\uppercase{Empathy}}
\label{sec:empathy}

Decety and Cowell \cite{Decety2014} say empathy is the ability to share in and understand others' experiences vicariously. There is broad consensus that empathy is a fundamental component of our social and emotional lives. Indeed, empathy has a vital role in social interaction, including understanding others' feelings and subjective psychological states. Empathy-related processes are thought to motivate prosocial behavior (e.g., sharing, comforting, and helping) and caring for others, inhibit aggression, and provide the foundation for care-based morality. 

Henschel et al. \cite{Henschel2020} say empathy corresponds to the ability to understand others' minds, feel their emotions outside our own, and respond with kindness, concern, and care to their emotions. It is a multidimensional construct encompassing an affective component (i.e., tendencies to feel compassion and concern for others) and a cognitive component (i.e., an ability to understand the reasons for another person's emotions and imagine different viewpoints beyond one's own). 

Cameron et al. \cite{Cameron2019} suggest empathy may not be easy — in many cases, particularly with strangers, it may require cognitive work. Empathy may seem less taxing for loved ones or in environments that scaffold empathy with social rewards and may be selected rather than suppressed. People may set the limits of empathy based on how hard they want to work. A study from Weisz and Zaki \cite{Weisz2018} suggests that people want to empathize with those most relevant to them. This tendency goes beyond group membership; people are motivated to empathize with those who look like them, those who are kind to them, and those close to them. 

Dave Gray created the empathy map technique in 2009 to help teams develop deep, shared understanding and empathy for other people. People use it to improve customer experience, navigate organizational politics, design better work environments, and host other things. The original goal of the empathy map is to gain a deeper level of understanding of a stakeholder in your business ecosystem, which may be a client, prospect, partner, etc., within a given context, such as a buying decision or an experience using a product or service \cite{Gray2017}. 

In Human-Computer Interaction (HCI), empathy appears in different works as in Bennett and Rosner \cite{Bennett2019} that studied empathy around disability. Gonzales et al. \cite{GonzalezBanales2017} used the empathy map as a tool to analyze Human-Computer Interaction in the elderly. Ferreira et al. \cite{Ferreira2015} used a combination of personas and empathy maps to enhance user experience.
In the context of Software Engineering, we opted to use the technique experimentally in a different situation, so we used the Empathy Map Canvas technique to collect the data through our survey. We consider it essential to connect with the participants' experiences and their emotions for this sensitive subject. It was a valid experience once two respondents mentioned that the survey touched them, and it got them thinking about their professional journey.

\begin{table*}[ht]
\centering
\caption{Demographics of the Participants of the Study}
\label{tab:demographics}
\resizebox{0.9\textwidth}{!}{%
\begin{tabular}{@{}ccccccccc@{}}
\toprule
    & Answer & Age  & Gender & \begin{tabular}[c]{@{}c@{}}Race/\\ Ethnicity\end{tabular} & \begin{tabular}[c]{@{}c@{}}Years\\ in IT\end{tabular} & \begin{tabular}[c]{@{}c@{}}Years in\\ current job\end{tabular} & Role                                                                                 & \begin{tabular}[c]{@{}c@{}}Self Evaluation \\ on experience\end{tabular} \\  \rowcolor{lightgray}\midrule
    & \#1    & 34   & Woman  & White                                                     & 10                                                    & 1                                                              & Scrum Master\ IT Manager                                                                         & \begin{tabular}[c]{@{}c@{}}Specialist\\ (+3 years)\end{tabular}          \\
    & \#2    & 43   & Woman  & White                                                     & 19                                                    & 2                                                              & \begin{tabular}[c]{@{}c@{}}ScrumMaster\\ Product Owner\\ People Manager\end{tabular} & \begin{tabular}[c]{@{}c@{}}Specialist\\ (+3 years)\end{tabular}          \\
   \rowcolor{lightgray} & \#3    & 35   & Woman  & White                                                     & 16                                                    & 2                                                              & IT Coordinator                                                                       & \begin{tabular}[c]{@{}c@{}}Experienced\\ (1 - 3 years)\end{tabular}      \\
\bottomrule
Avg &        & 37.3 &        &                                                           & 15                                                    & 1.7                                                            &                                                                                      &                                                                          \\ \bottomrule
\end{tabular}%
}
\end{table*}

\section{\uppercase{The Study Design}}
\label{sec:design}

Given \cite{Given2008} says survey research refers to the set of methods used to gather data systematically from a range of individuals, organizations, or other units of interest. Specific methods may include questionnaires (on paper or online), interviews (conducted by any method; e.g., individual interviews done face to face or via telephone), focus groups, or observation (e.g., structured observations of people using internet access stations at a public library).

This work's data collection began when people in Brazil were around four months in social distancing due to the COVID-19 pandemics. The original idea was to perform a focus group with three women in leadership/management roles using the empathy map canvas questions to guide the discussion. However, the pandemics forced us to change the plans, and we opted to adapt the empathy map canvas to an online survey. Two were the main reasons to send the empathy map to respond as a survey instead of running a focus group using online conference call tools: 
\begin{enumerate}
\item  Schedule: the professionals' agendas are busier than before COVID-19. People report that they have more meetings than usual, even after office hours, which leads them to exhaustion. 
\item People are reporting higher sensitivity to their emotions and difficulties in expressing them in a group.

\end{enumerate}
Given the situation, after a brief conversation with the women who were going to participate in the focus groups, we opted for sending the empathy map in an online survey format so they could respond asynchronously and in the most comfortable moment for them.

Given \cite{Given2008} says the defining element of focus groups is the use of the participants' discussion as a form of data collection. In particular, there is no requirement to reach a consensus or produce a decision; instead, it is the participants' conversation about the research topic of interest. Once, initially, the focus group would be guided by the empathy map canvas questions, we entirely mapped the questions as open-questions to the online survey. Given \cite{Given2008} also says open-ended questions provide greater freedom to the researcher in terms of how to frame the question, as well as granting greater freedom to respondents in the ways they choose to answer.  Open-ended questions may challenge respondents because they are more demanding and time-consuming to answer; however, the data obtained are typically richer than that generated from closed questions. Considering this, we understood that we could lose the conversation between the participants, but we still would have rich qualitative data to analyze.

Below, we list the high-level questions of the empathy map as defined by Dave Gray \cite{Gray2017}, and we included in the survey:

\begin{enumerate}
\item Who we are empathizing with?
\begin{itemize}
\item  Who is the person we want to understand? 
\item  What is the situation they are in? 
\item  What is their role in the situation? 
\item  How will we know they were successful?
\end{itemize}

\item What do they need to do?  
\begin{itemize}
\item What do they need to do differently? 
\item  What tasks do they want or need to get done? 
\item  What decisions do they need to make?
\end{itemize}
        
\item What do they see?   
\begin{itemize}
\item  What are they watching and reading? 
\item  What kind of tasks are they exposed to daily? 
\item  Who are their friends? 
\item  What kind of problems they face?  
\end{itemize}
  
\item What do they say?
\begin{itemize}
\item What have we heard they say? 
\item  What can we imagine they saying?
\end{itemize}

\item What do they do?
\begin{itemize}
\item What do they do today? 
\item  What behavior have we observed? 
\item  Whats can we imagined they doing?
\end{itemize}
    
\item What do they hear?
\begin{itemize}
\item What are they hearing others saying? 
\item  What are they hearing from friends? 
\item  What are they hearing from colleagues? 
\item  What do they hear second hand?
\end{itemize}

\item What do they think and feel?
\begin{itemize}
\item Pains: What are their fears, frustrations, and anxieties?
\item Gains: What are their wants, needs, hopes, and dreams?
\end{itemize}

\end{enumerate}
\subsection{Participants}
We invited three women in Software Engineering leadership/management roles to answer our questions.  They work in the same tech company in Brazil, but they have different career paths. They have worked for different companies and have different work and life experiences. Table \ref{tab:demographics} presents the demographic data of those women. We did not provide pre-filled gender and race/ethnicity lists, so respondents could self-identify themselves.

\subsection{Qualitative Data Analysis}

Runeson and Martin \cite{Runeson2009} say analysis of qualitative data is conducted in a series of steps. First, the data is coded, which means that parts of the text can be given a code representing a certain theme, area, construct, etc. One code is usually assigned to many pieces of text, and one piece of text can be assigned more than one code. Codes can form a hierarchy of codes and sub-codes. The coded material can be combined with the researcher's comments and reflections (i.e., "memos"). When this has been done, the researcher can go through the material to identify a first set of hypotheses. For example, these can be phrases similar in different parts of the material, patterns in the data, differences between sub-groups of subjects, etc. The identified hypotheses can then be used when further data collection is conducted in the field, i.e., resulting in an iterative approach where data collection and analysis is conducted in parallel as described above. During the iterative process, a small set of generalizations can be formulated, eventually resulting in a formalized body of knowledge, which is the final result of the research attempt. This is, of course, not a simple sequence of steps. Instead, they are executed iteratively, and they affect each other.

We performed a thematic analysis with open coding on the responses from the empathy map. We read and reread the data, looking for keywords, trends, themes, or ideas to outline the analysis. This analysis identified themes across the questions from the survey. Table \ref{tab:thematic} lists the codes, themes, and quotes. 


\begin{table*}[]
\centering
\caption{Results of the Thematic Analysis}
\label{tab:thematic}
\resizebox{0.9\textwidth}{!}{%
\begin{tabular}{|l|l|l|}
\hline
{\color[HTML]{0E101A} \textbf{Codes}}                                                                                      & {\color[HTML]{0E101A} \textbf{Theme}}                                                                             & {\color[HTML]{0E101A} \textbf{Quotes}}                                                                                                                                                                                                                                                                                                                                                                                                                                                                                                                                                                                                                                                                                                                                                                                                        \\ \hline
{\color[HTML]{0E101A} \begin{tabular}[c]{@{}l@{}}Authorization\\ Deserve\\ Right\\ Positions\end{tabular}}                 & {\color[HTML]{0E101A} \textbf{\begin{tabular}[c]{@{}l@{}}Permission to Occupy \\ Technology Spaces\end{tabular}}} & {\color[HTML]{0E101A} \textit{\begin{tabular}[c]{@{}l@{}}“I see in the company I  work, an exact sample of the market: \\ women occupying “authorized spaces.” It seems that I have been \\ given the right to be where I am and that it is the part I deserve."\\ \\ “The deconstruction of the thought that women are not able \\ to occupy positions that men predominantly occupy today."\end{tabular}}}                                                                                                                                                                                                                                                                                                                                                                                                                                  \\ \hline
{\color[HTML]{0E101A} \begin{tabular}[c]{@{}l@{}}Recognition\\ Opportunities\\ Equity\end{tabular}}                        & {\color[HTML]{0E101A} \textbf{\begin{tabular}[c]{@{}l@{}}Recognition and \\ Equal Opportunities\end{tabular}}}    & {\color[HTML]{0E101A} \textit{\begin{tabular}[c]{@{}l@{}}“Recognition and equity in opportunities.”\\ \\ “Recognition and opportunity to have more space.”\\ \\ “A plural and psychologically healthy professional \\ environment that provides equity in career development”\end{tabular}}}                                                                                                                                                                                                                                                                                                                                                                                                                                                                                                                                                  \\ \hline
{\color[HTML]{0E101A} \begin{tabular}[c]{@{}l@{}}Beyond\\ Potential\end{tabular}}                                          & {\color[HTML]{0E101A} \textbf{\begin{tabular}[c]{@{}l@{}}Need to Go \\ Above and beyond\end{tabular}}}            & {\color[HTML]{0E101A} \textit{\begin{tabular}[c]{@{}l@{}}“I often feel that I need to go beyond my peers in terms of training as if \\ I was never competent enough to achieve the results expected from me”\\ \\ “I still see something very distant from diverse team structures. \\ I already see some diversity at the beginning of career. In development teams, \\ I don’t see a woman Specialist, for example.”\\ \\ “At the beginning of my career, I always thought that when I was 40 years old, \\ I would be someone in a strategic area that would make a difference in a scaled way. \\ I know it makes a difference, but it is not a universe in scale. \\ I often see that my potential is not used at their maximum.”\end{tabular}}}                                                                                         \\ \hline
{\color[HTML]{0E101A} \begin{tabular}[c]{@{}l@{}}Tiredness\\ Expectations\\ Prove\\ Judgement\end{tabular}}                & {\color[HTML]{0E101A} \textbf{\begin{tabular}[c]{@{}l@{}}Mental wear \\ and tear\end{tabular}}}                   & {\color[HTML]{0E101A} \textit{\begin{tabular}[c]{@{}l@{}}“I feel tired because I want to make a difference, so \\ I put a lot of effort into thinking and planning before proposing something.”\\ \\ I feel they expect I need to have always a position about everything.\\  I can not exempt myself from an opinion.”\\ \\ “I would say that the biggest problem is the counteraction. \\ I always need to present hard proof of my suggestions and plans.”\\ \\ “Frequently, mental wear and tear for performing tasks in very different contexts.”\\ \\ “As a woman and mother, I need to prove myself much more. \\ If a man and a woman perform the same task and are successful, \\ the man will be praised and promoted for it, \\ while the woman will be criticized, and her outcome will be judged with suspicion.”\end{tabular}}} \\ \hline
{\color[HTML]{0E101A} \begin{tabular}[c]{@{}l@{}}Drama\\ Crazy\end{tabular}}                                               & {\color[HTML]{0E101A} \textbf{\begin{tabular}[c]{@{}l@{}}Drama and \\ Stereotypes\end{tabular}}}                  & {\color[HTML]{0E101A} \textit{\begin{tabular}[c]{@{}l@{}}“I also imagine many teams saying that “everything is drama.”\\ \\“Do I need to change my way of speaking and acting with that person then?” \\ “Women are dramatic and talk too much,” \\ \\“I don’t know how to talk about it, I prefer not to talk,” \\ \\“Women with short hair are not women,” \\ \\“That woman is crazy, “etc.\end{tabular}}}                                                                                                                                                                                                                                                                                                                                                                                                                                          \\ \hline
{\color[HTML]{0E101A} \begin{tabular}[c]{@{}l@{}}Male Protectionism\\ Masculinization\\ Sexism\\ Humiliation\end{tabular}} & {\color[HTML]{0E101A} \textbf{\begin{tabular}[c]{@{}l@{}}Obstacles \\ and Fears\end{tabular}}}                    & {\color[HTML]{0E101A} \textit{\begin{tabular}[c]{@{}l@{}}“A network of male protectionism, which prevents women from advancing \\ to strategic positions (or advancing with great difficulty, requiring \\ a certain masculinization for that).”\\ \\ “the partnerships that count a lot at all levels. Do not be a white man.”\\ \\ “Structural sexism and even to be scolded for trying to fight against it.”\\ \\ “To stay out of the standard behavior that companies expect from \\ the role that I find myself in today (software engineering manager) \\ for the sake of being myself.”\\ \\ “Once again humiliation”\end{tabular}}}                                                                                                                                                                                                   \\ \hline
{\color[HTML]{0E101A} \begin{tabular}[c]{@{}l@{}}Hiring\\ Diversity\\ Education\\ Make Difference\end{tabular}}            & {\color[HTML]{0E101A} \textbf{\begin{tabular}[c]{@{}l@{}}Hire and Education\\ for Diversity\end{tabular}}}        & {\color[HTML]{0E101A} \textit{\begin{tabular}[c]{@{}l@{}}“I carry out actions to expand the hiring of women and black people. \\ It is also my goal to carry out education actions with my teams”\\ \\ “Once I am the only woman in management in my area, the behavior \\ I seek is not to follow the established pattern. I seek to bring diversity to my teams.”\\ \\ “When hiring, I seek diversity and bring diversity. \\ I have a fortnightly meeting group with this diverse group where \\ we seek a supportive and safe environment. But my goal behind \\ this action is beyond that I want this diverse group to be the \\ next generation of leadership. At the base, it is challenging \\ to make the difference that needs to be made.”\end{tabular}}}                                                                         \\ \hline
\end{tabular}%
}
\end{table*}

\section{\uppercase{Results and Discussion}}\label{sec:discussion}

The data collected through the empathy map applied to three software engineering managers were coded and grouped in themes, as shown in Table \ref{tab:thematic}. We identified 23 codes then we grouped them into seven themes. The themes helped us answer the research question proposed for this study, and we discuss them in this Section.

\subsection{RQ. What are the challenges women in Software Engineering middle management roles face?}

\textbf{Permission to Occupy Authorized Spaces}
As mentioned at the beginning of this work, Frize \cite{Frize2005} says there are serious gender issues that can severely limit women's participation in science and engineering careers, and these are similar in many parts of the world. One main obstacle to women's retention or participation is that women's contributions and abilities are less valued than men's, and women are generally ignored in mainstream history.  From the answers we collected, it is possible to see that women want to occupy positions that men traditionally occupied. However, there is also a sensation that women can occupy those roles when "authorized." The positions occupied for those women are as they were given the right to be there, and it is what they deserve. 

\textbf{Recognition and Equal Opportunities}
The three managers reported their desire for recognition.  They want to know that what they are doing is significant and acknowledge by their peers and managers. The recognition comes together with the desire for equal opportunities to grow in their career as their men peers.

\textbf{Need to Go Above and Beyond}
The managers mentioned they feel like they always need to put more effort and go beyond the men's peers. There is the sensation that the same work is praised when a man does it, and it is suspicious when a woman does. Again, women mention the high effort to prove their work is good or even better than their male peers' work. 

\textbf{Mental Wear and Tear}
The efforts for recognition and the high energy put into achieving it leads them to exhaustion. The managers mentioned they feel tired because they always need to have a strong opinion and a solution for every problem shared with them.

\textbf{Drama and Stereotypes}
Eagly and Carli \cite{EAGLY2003} say prejudice consists of an unfair evaluation of a group of people based on stereotypical judgments of the group rather than the behavior or qualifications of its members. When people hold stereotypes about a group, they expect that group to possess characteristics and exhibit behavior consistent with those stereotypes. In our study, the surveyed women mentioned stereotypes about their physical appearance. They said they are judged by what they wear, by the makeup absence, and even for have opted for short hair ("\textit{women with short hair are not women}"). Also, their mental state is always put in check. When sharing concerns about keeping a healthy and psychologically safe environment, they use to hear that look for it is an exaggeration and that they are doing "\textit{drama}." They also listen to things as "\textit{this woman is crazy}."

\textbf{Obstacles and Fears}
The managers mentioned structural sexism in different parts of the survey. Sometimes they are afraid of talking about it once their male peers or even their managers can scold them. As a considerable obstacle, they see the brotherhood between men at all levels. They mentioned the perception of the existence of "\textit{a network of male protectionism, which prevents women from advancing to strategic positions}." 

\textbf{Hire and Education for Diversity}
Besides the heavy psychological load women carry out to conduct their daily activities, they are still concerned about creating psychological safety for gender diversity. As hiring managers, they use the opportunity to build more diverse and inclusive environments in technology. With that, they try to minimize the challenges to the next generations of women in leadership positions. However, they report low support from peers and higher levels.\\

Besides the challenges, the three managers mentioned that even with the small representativeness, they believe in paving the way for more women in technology careers. They are supported by thoughts of doing something meaningful to other people's lives and changing the picture, even with small acts.

Additionally to the results, we considered using the empathy map in this study as a valid experience. Two respondents mentioned that the survey touched them deeply, and it got them thinking about their professional journey after answering.

\subsection{Threats to Validity}

Any empirical study is subject to several threats to validity. This section is organized by classification of the threats to validity in three classes: Internal, External, and Construct categories \cite{wohlin2012}

\textbf{Internal validity} is the extent to which the design and conduct of the study are likely to prevent systematic error \cite{kitchenham2007}. It concerns confounding factors that can influence the obtained results. We assumed a causal relationship between the perceptions through the entire set of answers we coded and grouped in themes and the answers with what is considered challenging for the women that answered the survey.

 \textbf{External validity} is the extent to which the effects observed in the study are applicable outside of the study \cite{kitchenham2007}. The presented results are related to answers of three women in roles of software engineering management working for the same company, so the results are only valid in this context, and we cannot generalize them. However, this work does not intend to be a complete analysis of women's situation in software engineering management roles. It is only the first step of a study using the empathy map to collect qualitative data. An extension of this work is needed to confirm our findings.

Threats to \textbf{Construct validity} focus on how accurately the observations describe the phenomena of interest. The coding process is subject to researcher bias, as the process of grouping the codes in themes. To minimize bias due to personal interpretation, the process was reviewed by a second researcher.

\section{\uppercase{Conclusion}}
\label{sec:conclusion}
This work does not intend to be a complete analysis of women's situation in software engineering middle management roles. We also cannot generalize the results, and no intersectionality evaluation could be done once the three women reported being white women. Intersectionality is the theoretical framework for understanding how aspects of a person's social and political identities (e.g., gender, sex, race, class, sexuality, religion, ability, physical appearance, height, etc.) might combine to create unique modes of discrimination and privilege, coined by Kimberle Crenshaw \cite{Kimberle1989}. However, we intended to bring some initial insights from applying the empathy map technique to these women and raise awareness of what impacts their careers, the challenges they face, and show the distress they are exposed to. This work is the initial step of a broader work where using empirical strategies; we aimed to understand if a software development team with greater gender diversity performs better than a homogeneous team and which factors may be the most impacted by this spectrum of diversity. Data collection is happening in different companies from the information technology industry through surveys, interviews, and code repository analysis from other software development teams' roles. With that, we expect to reach a broader number of women of different races/ethnicities, cultures, etc. So we will be able to analyze the data from the point of view of intersectionality.

\section*{\uppercase{Acknowledgements}}
This project is partially funded by FAPERGS, project 17/2551-0001/205-4.

\bibliographystyle{apalike}
{\small
\bibliography{example}}

\begin{thebibliography}{}

\bibitem[Bennett and Rosner, 2019]{Bennett2019}
Bennett, C.~L. and Rosner, D.~K. (2019).
\newblock {The promise of empathy: Design, disability, and knowing the
  “other”}.
\newblock {\em Conference on Human Factors in Computing Systems - Proceedings},
  pages 1--13.

\bibitem[Cameron et~al., 2019]{Cameron2019}
Cameron, C., Hutcherson, C., Ferguson, A., Scheffer, J., Hadjiandreou, E., and
  Inzlicht, M. (2019).
\newblock Empathy is hard work: People choose to avoid empathy because of its
  cognitive costs.
\newblock {\em Journal of Experimental Psychology: General}, 148(6):962--976.

\bibitem[Crenshaw, 1989]{Kimberle1989}
Crenshaw, K. (1989).
\newblock Demarginalizing the intersection of race and sex: A black feminist
  critique of antidiscrimination doctrine, feminist theory and antiracist
  politics.
\newblock {\em University of Chicago Legal Forum}, 1989(3):207--236.

\bibitem[Decety and Cowell, 2014]{Decety2014}
Decety, J. and Cowell, J.~M. (2014).
\newblock {The complex relation between morality and empathy}.
\newblock {\em Trends in Cognitive Sciences}, 18(7):337--339.

\bibitem[Eagly and Carli, 2003]{EAGLY2003}
Eagly, A.~H. and Carli, L.~L. (2003).
\newblock The female leadership advantage: An evaluation of the evidence.
\newblock {\em The Leadership Quarterly}, 14(6):807 -- 834.

\bibitem[Ferreira et~al., 2015]{Ferreira2015}
Ferreira, B., Conte, T., and Barbosa, S. D.~J. (2015).
\newblock {Eliciting Requirements Using Personas and Empathy Map to Enhance the
  User Experience}.
\newblock {\em Proceedings - 29th Brazilian Symposium on Software Engineering,
  SBES 2015}, pages 80--89.

\bibitem[Frize, 2005]{Frize2005}
Frize, M. (2005).
\newblock Women in leadership: Value of women's contributions in science,
  engineering, and technology.
\newblock In {\em Proceedings of the International Symposium on Women and ICT:
  Creating Global Transformation}, CWIT '05, page 4–es, New York, NY, USA.
  Association for Computing Machinery.

\bibitem[Given, 2008]{Given2008}
Given, L.~M. (2008).
\newblock {\em Qualitative Research Methods - Volumes 1-2. The SAGE
  Encyclopedia of Qualitative Research methods.}

\bibitem[Gonz{\'{a}}lez-Ba{\~{n}}ales and Ort{\'{i}}z,
  2017]{GonzalezBanales2017}
Gonz{\'{a}}lez-Ba{\~{n}}ales, D.~L. and Ort{\'{i}}z, L. E.~S. (2017).
\newblock {Empathy map as a tool to analyze human-computer interaction in the
  elderly}.
\newblock {\em ACM International Conference Proceeding Series}.

\bibitem["Gray, 2017]{Gray2017}
"Gray, D. (2017).
\newblock "update to the empathy map".

\bibitem[Henschel et~al., 2020]{Henschel2020}
Henschel, S., Nandrino, J.-L., and Doba, K. (2020).
\newblock Emotion regulation and empathic abilities in young adults: The role
  of attachment styles.
\newblock {\em Personality and Individual Differences}, 156:109763.

\bibitem[{Jetter} et~al., 2013]{Jetter2013}
{Jetter}, A.~J., {Loanzon}, E., {Jahromi}, S., {Nour}, A.~H., and
  {Pakdeekasem}, P. (2013).
\newblock An exploratory study on the leadership style preferences of male and
  female managers: Implications on team performance.
\newblock In {\em 2013 Proceedings of PICMET '13: Technology Management in the
  IT-Driven Services (PICMET)}, pages 1161--1181.

\bibitem[Jetter and Walker, 2017]{Jetter2017}
Jetter, M. and Walker, J.~K. (2017).
\newblock {The gender of opponents: Explaining gender differences in
  performance and risk-taking?}
\newblock {\em European Economic Review}, 109:238--256.

\bibitem[Kitchenham and Charters, 2007]{kitchenham2007}
Kitchenham, B.~A. and Charters, S. (2007).
\newblock Guidelines for performing systematic literature reviews in software
  engineering.
\newblock Technical Report EBSE 2007-001, Keele University and Durham
  University Joint Report.

\bibitem[Novielli and Serebrenik, 2019]{Novielli2019}
Novielli, N. and Serebrenik, A. (2019).
\newblock Sentiment and emotion in software engineering.
\newblock {\em IEEE Software}, 36(5):6--9+23.

\bibitem[Page, 2007]{Page2007}
Page, S.~E. (2007).
\newblock {\em The Difference: How the Power of Diversity Creates Better
  Groups, Firms, Schools, and Societies}.
\newblock Princeton University Press.

\bibitem[Runeson and H{\"{o}}st, 2009]{Runeson2009}
Runeson, P. and H{\"{o}}st, M. (2009).
\newblock {Guidelines for conducting and reporting case study research in
  software engineering}.
\newblock {\em Empirical Software Engineering}, 14(2):131--164.

\bibitem[Sytsma, 2006]{Sytsma2006}
Sytsma, S.~E. (2006).
\newblock {\em Ethics and intersex}.
\newblock (Springer e-books.) Dordrecht: Springer.

\bibitem[Usher, 2006]{usher2006}
Usher, R. (2006).
\newblock {\em North American Lexicon of Transgender Terms.}
\newblock GLB Publishers.

\bibitem["Weisz and Zaki, 2018]{Weisz2018}
"Weisz, E. and Zaki, J. (2018).
\newblock "motivated empathy: a social neuroscience perspective".
\newblock {\em Current opinion in psychology}, 24:67–71.

\bibitem[Wohlin et~al., 2012]{wohlin2012}
Wohlin, C., von Mayrhauser, A., Runeson, P., H{\"o}st, M., Ohlsson, M.,
  Regnell, B., and Wessl{\'e}n, A. (2012).
\newblock {\em Experimentation in Software Engineering: An Introduction}.
\newblock International Series in Software Engineering. Springer US.

\bibitem[Wohlin~C., 2003]{Wohlin2003}
Wohlin~C., Höst~M., H.~K. (2003).
\newblock Empirical research methods in software engineering.
\newblock In Conradi~R., W.~A., editor, {\em Lecture Notes in Computer
  Science}, volume 2765, pages 266--290. Springer, Berlin, Heidelbergs.

\end{thebibliography}

\end{document}